%% file: 0_main.tex
\documentclass[sigconf,nonacm]{acmart}
\AtBeginDocument{%
  }

\acmConference[CSCW '26 Workshop]{Broader Impacts of GenAI in Communication}{October, 2026}{Salt Lake City, UT}

\begin{document}

\title[How Much Can AI Understand?]{How Much Can AI Understand? Toward AI-Assisted Sensemaking of Collaborative Discussion in Groups with Shared History}


\author{Soobin Cho}
\email{soobin30@uw.edu}
\affiliation{%
  \institution{Human Centered Design \& Engineering, University of Washington}
  \city{Seattle}
  \country{USA}
}

\author{Mark Zachry}
\email{zachry@uw.edu}
\affiliation{%
  \institution{Human Centered Design \& Engineering, University of Washington}
  \city{Seattle}
  \country{USA}
}

\author{David W. McDonald}
\email{dwmc@uw.edu}
\affiliation{%
  \institution{Human Centered Design \& Engineering, University of Washington}
  \city{Seattle}
  \country{USA}
}

\renewcommand{\shortauthors}{Cho et al.}

\begin{abstract}
AI tools that support collaborative discussion typically treat the discussion as a standalone task, focusing only on its content and setting aside the social context of the group having it. But it is groups with a shared history, with their own norms, hierarchies, and relationships, where the most tangled and complex discussions tend to arise. These discussions cannot be understood apart from that context, and AI that overlooks it risks failing to convey what a discussion means, or even misrepresenting it. Drawing on two studies of how experienced Wikipedia editors read and make sense of discussions, we propose an AI-Assisted Sensemaking Model for Collaborative Discussions, which captures not only a discussion's arguments but also the norms and participants behind it, along with the context that gives each meaning. In this model, the system supports the early stages of the sensemaking process, and the degree to which it performs interpretive work can range from low to high. We argue that higher interpretive work reduces the burden on users but increases their reliance on the system's judgment. We then discuss the risks of an insufficiently intelligible system, what it would take to make one more intelligible, and the safeguards it still requires.
\end{abstract}


\maketitle

\section{Motivation}
Discussion is essential to how collaborative working groups get work done. It is how members understand each other's positions, work through disagreement, and reach the decisions needed to move forward. Some discussions are short and resolve quickly, but just as many stretch on and grow complex, which can mean becoming entangled with earlier practices, or having the people involved shift over time. Anyone who has been part of a long-running collaborative working group, whether a project team at a company, a research team at a lab, or a volunteer team, has likely had to make sense of a discussion that grew complex because of a myriad of factors.

CSCW has long studied ways to support discussion in collaborative settings, with a large body of work on tools such as argument visualization~\cite{kunz1970issues, conklin1988gibis} and group decision support tools~\cite{desanctis1987foundation, Nunamaker}. As AI has advanced, research on autonomous systems that support discussion has grown as well. This includes argument mining~\cite{palau2009argumentation, lawrence2020argument}, which automatically extracts argumentative structure from text, and discussion facilitation bots~\cite{Kim1, Kim2, Do1}, which are conversational agents that take on the role of a discussion facilitator. With the rise of LLMs, practical applications now include agents that generate meeting minutes, while research has investigated how to generate consensus statements that participants in public deliberation can agree on \cite{Tessler}.

However, these AI systems typically focus only on the given discussion content, without considering the social context around it. Research on these systems also tends to assume an ad~hoc group formed only for the task at hand. But it is collaborative working groups with their own history where the most complex discussions tend to arise, and these groups' discussions cannot be separated from their social context. An AI that attempts to assist discussion without taking this social context into consideration risks failing to fully grasp and convey the meaning that contributions within the discussion carry, or even misrepresenting them.

Groups with a shared history usually have explicit or implicit ways of working and communicating, and members differ in their experience, positions within the group's hierarchy, and relationships with one another, all of which are naturally reflected in how a discussion unfolds. In a project team, for instance, members may know that a given argument belongs to an important stakeholder and that the group tends to defer to it. A new member joining the project would find little help in an AI that summarizes only the arguments themselves, and it would probably take a senior member quietly filling in that context before the newcomer could truly make sense of the discussion.

This raises the question of how sensemaking of discussion unfolds in collaborative working groups with a shared history, and how AI can support it.

\section{AI-Assisted Sensemaking Model for Collaborative Discussions}

\input{model}

Our understanding of how members of a collaborative working group with a shared history make sense of discussions derives from studies we conducted in the context of Wikipedia. In Wikipedia, contributors work together to evaluate content and build consensus on what should be included in encyclopedic articles. To carry out this collaborative work, editors engage in a large number of discussions, many of which run long and grow complex over time. Wikipedia is governed by an extensive and interlinked system of community-created policies and guidelines~\cite{wiki:norms}. However elaborate these are, they are applied in practice with their own implicit conventions, some of which are informally documented in a space called essays~\cite{wiki:essay}, though these essays inevitably fall short of capturing every convention. Registered editors' identities are reflected in their edit histories and user pages, which convey their Wikipedia experience and background, as well as in their interactions with other editors across various pages, which reflect their past relationships and social roles~\cite{Cosley}.

To understand how experienced Wikipedians read and make sense of complex discussions, we conducted two studies. The first examined their cognitive process as they read discussions unassisted, identifying the information elements they attend to~\cite{cho}. The second developed a web-based discussion sensemaking prototype with LLM-assisted features and studied how Wikipedians used it, along with the cognitive process behind their use (under review). Based on these two studies, we developed an AI-assisted sensemaking model for collaborative discussions (Figure~\ref{fig:model}).

Through our study, we found that sensemaking operates over six types of information: argument, community norm, norm context, people, people context, and discussion topic. While argument, community norm, and people are directly expressed within the discussion, norm context and people context lie outside the discussion within the broader community, and discussion topic represents domain-level information that exists outside the community.

We found that participants' sensemaking process can be aligned with Pirolli and Card's sensemaking model~\cite{Pirolli&Card}, so the human part of our model follows their stages: Search \& Filter $\rightarrow$ Read \& Extract $\rightarrow$ Schematize $\rightarrow$ Build Case $\rightarrow$ Tell Story. In their model, sensemaking begins with Search and Filter, which identifies and collects relevant information, followed by Read and Extract, which pulls useful evidence from what was collected. This evidence is then Schematized into a structured representation, Build Case develops an interpretation from that representation, and Tell Story communicates the resulting interpretation. Inquiries or feedback on the interpretation can direct attention back to earlier stages, prompting re-evaluation and further search.

The AI system in our model operates as an overlay on the early stages of the human process. It performs four functions: Extraction, Prioritization, Connecting, and Presentation. Extraction involves identifying and extracting information elements from raw data; Prioritization involves selecting and ordering information based on importance; Connecting establishes relationships between different information elements; and Presentation provides these results as intermediate outputs. Prioritization and Connecting can operate iteratively, repeatedly informing one another as importance is reassessed based on newly formed relationships.

In our model, the degree of interpretive work the system performs can range from low to high. We define interpretive work as how much the system modifies raw data, based on the importance of that information for final interpretation. For every individual information feature, such as a discussion summary, an overview of participant information, or past norm practices, designers need to decide on the degree of its interpretive work. For example, in presenting a discussion summary, one lower-interpretation implementation might preserve much of the original discussion while only lightly condensing it. As the system produces a more selective summary that includes only the points it identifies as important, it performs more interpretive work.

\section{Difficulty of Interpreting Discussion Correctly, and What It Means for System Design}
Any time an AI system, an LLM or otherwise, is applied to a task, it performs interpretive work, whether or not its designers or users notice. This is not unique to our context or model. However, that work is not always correct. Hallucination is one example of what an incorrect result can look like. In our context of collaborative discussion, being correct means correctly interpreting participants' arguments during discussion sensemaking.

\subsection{Interpreting Human}
To correctly interpret someone's argument is to accurately understand the meaning behind what they said. This is essentially what Max Weber~\cite{weber1981some} defines as the goal of interpretive sociology, an attempt to ``understand'' human behavior through ``interpretation.'' According to Weber, this process aims to arrive at an ``intelligible explanation'' of that behavior.

Reaching this goal requires keeping human subjectivity in mind. The behavior interpretive sociology seeks to understand is grounded in the actor's own subjectively intended meaning, and this behavior is interrelated with and codetermined by others' behavior. In other words, one's action is shaped by how the actor subjectively understands the meaning of their own action in relation to others. The same subjective logic applies to ``associational action,'' an action oriented toward an established rule or order that applies to the whole group, the actor along with the other actors. The order itself exists objectively, but each actor takes it into account subjectively, and how each actor understands and expects others to follow that order is a subjective judgment.

Irrationality is another feature of social action Weber discusses. He warns that if interpretive analysis only assumes rationality, the researcher will keep running into purposes that can no longer be interpreted as rational means to some further end. Irrational action is diverse, ranging from irrational but understandable behavior to behavior that resists understanding altogether. Weber also notes that even an action starting from the same rational basis can, through incidental factors, take an irrational course.

Underlying both features, subjectivity and irrationality, is emotion. Weber treats emotion as not quite action itself, but closely bound to social action. It shapes how an actor subjectively constructs their action's intended meaning, and can be a source of irrational action.

Weber's account shows just how difficult correctly interpreting someone's behavior is, given human subjectivity and irrationality, and the role emotion plays in both.

\subsection{Implications for the System's Interpretive Work}
We distinguish between interpretation and interpretive work. Interpretive work is the process and practice through which interpretation happens, while interpretation is the meaning ultimately arrived at through that process. In our model, the system performs interpretive work rather than interpretation, filtering and highlighting the information that matters for the human's eventual interpretation. Yet even though the system does not interpret participants' arguments directly, its interpretive work still needs to be ``intelligible'', especially given the complexity of the social setting a discussion like this involves. Whether a system is capable of that is an open question.

\subsubsection{The Risks of a System That Falls Short of Intelligibility}
Why this capability matters is that the level of interpretive work the system performs, low or high, corresponds inversely to the amount of interpretive work users need to do themselves. As the system's interpretive work increases and data is reshaped around the information it judges to be important, the work users need to do correspondingly decreases. In the discussion summary example above, a more selective summary means users spend less time reading through details to find what matters. But reducing users' interpretive work also means relying more on system-generated judgments. In other words, while more interpretive outputs can offer more useful support for sensemaking, they also require users to trust the system's judgment more.

If a system performing such a high level of interpretive work fails to correctly identify what information matters for making sense of a discussion in a complex social setting, several risks can arise.

One is that it dehumanizes participants, flattening them into generic contributors and stripping away the context that explains why their contribution matters. What's worse, this loss does not happen randomly; it happens in a systematically biased way. Whose contribution is recognized ends up depending less on how important what someone actually said is, and more on how legible their contribution is to the system. This can steer the discussion in the wrong direction and lead to poor decision making.

Another is homogenization, or false consensus. By surfacing only the points it judges important, a system can erase minority opinions and nuance that were actually present in the discussion, points that may have genuinely mattered but are never taken into account. It can also make it look like a cleaner agreement was reached than actually existed, letting the group act on a decision that does not have the buy-in it appears to have, with the underlying disagreement potentially resurfacing as conflict later on.

\subsubsection{Designing for Intelligibility}
For a system to gain the ability to judge what information matters, it has to be an ``intelligible actor'' that understands the specific social setting. We argue that the most important step toward this is providing a clear account of what kinds of information matter for understanding that social setting. In our context of collaborative discussion, this means, as our model shows, not focusing only on the arguments themselves, but also making clear that the norms mentioned in the discussion and their norm context matter, as do the participants and their people context. Each of these, norm context and people context, in turn requires its own layered understanding. Norm context, for instance, may need to be understood at three levels, an objective description of the norm, how each member subjectively understands that objective rule, and, since disagreement over the norm likely stems from differences in this subjective understanding, what shared understanding the members eventually reached and enacted through discussion. People context is more complex still, requiring an understanding of each member's experience, background, role, relationships, and character, among other things.

But does doing this actually make the system a sufficiently intelligible actor? We cannot know for certain. This is why we still need safeguards against the risks of a system that falls short. One safeguard is leaving the final interpretation to humans. This is why, in our model, the system operates only as an overlay on the earlier stages of sensemaking, information search and organization, while leaving Build Case and Tell Story to the human. Another safeguard could be continuing to expose the original data, for transparency and explainability. This could be done by linking any presentation involving interpretive work back to the original data and providing a rationale for the interpretive work performed.

\section{Toward a Design Agenda}
As AI already takes on a growing role in tools people use for group discussion, such as meeting-minute agents, this raises open questions for design. We see these as falling into three parts, deciding the degree of interpretive work each system feature should perform, making the system itself a more intelligible actor, and building in the safeguards it still requires. We hope this paper offers a starting point for addressing them.

\bibliographystyle{ACM-Reference-Format}
\bibliography{references}

\end{document}

%% file: model.tex
\begin{figure*}[t]
    \centering
    \includegraphics[width=0.8\textwidth]{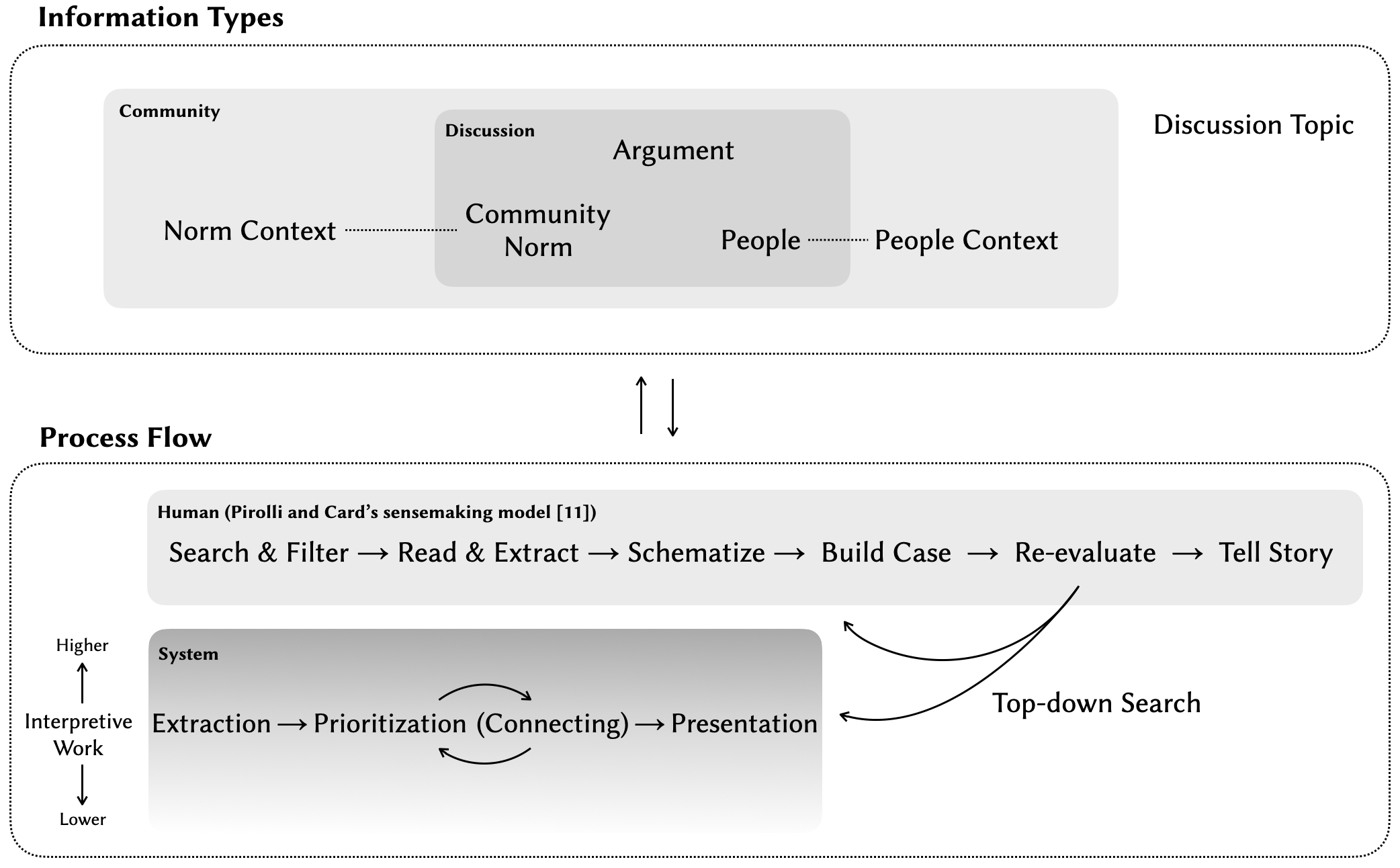}
    \caption{AI-assisted sensemaking model for collaborative discussions. Information types include argument, community norm, norm context, people, people context, and discussion topic. The model retains Pirolli and Card’s sensemaking model~\cite{Pirolli&Card} as the human component and extends it by introducing system support over the early stages of the human sensemaking process. The system performs Extraction, Prioritization, Connecting, and Presentation, operating along a spectrum from lower to higher system interpretive work.}
    \label{fig:model}
\end{figure*}

%% file: references.bib
@article{kunz1970issues,
  title={Issues as elements of information systems},
  author={Kunz, Werner and Rittel, Horst WJ},
  year={1970}
}

@article{conklin1988gibis,
  title={gIBIS: A hypertext tool for exploratory policy discussion},
  author={Conklin, Jeff and Begeman, Michael L},
  journal={ACM Transactions on Information Systems (TOIS)},
  volume={6},
  number={4},
  pages={303--331},
  year={1988},
  publisher={ACM New York, NY, USA}
}

@article{desanctis1987foundation,
  title={A foundation for the study of group decision support systems},
  author={DeSanctis, Gerardine and Gallupe, R Brent},
  journal={Management science},
  volume={33},
  number={5},
  pages={589--609},
  year={1987},
  publisher={INFORMS}
}

@article{Nunamaker,
author = {Nunamaker, Jay and Dennis, Alan and Valacich, Joseph and Vogel, Doug and George, Joey},
year = {1991},
month = {01},
pages = {40-61},
title = {Electronic Meeting Systems To Support Group Work.},
volume = {34},
journal = {Commun. ACM}
}

@article{lawrence2020argument,
  title={Argument mining: A survey},
  author={Lawrence, John and Reed, Chris},
  journal={Computational linguistics},
  volume={45},
  number={4},
  pages={765--818},
  year={2020},
  publisher={MIT Press One Rogers Street, Cambridge, MA 02142-1209, USA journals-info~…}
}

@inproceedings{palau2009argumentation,
  title={Argumentation mining: the detection, classification and structure of arguments in text},
  author={Palau, Raquel Mochales and Moens, Marie-Francine},
  booktitle={Proceedings of the 12th international conference on artificial intelligence and law},
  pages={98--107},
  year={2009}
}

@article{Do1,
author = {Do, Hyo Jin and Kong, Ha-Kyung and Tetali, Pooja and Karahalios, Karrie and Bailey, Brian P.},
title = {Inform, Explain, or Control: Techniques to Adjust End-User Performance Expectations for a Conversational Agent Facilitating Group Chat Discussions},
year = {2023},
issue_date = {October 2023},
publisher = {Association for Computing Machinery},
address = {New York, NY, USA},
volume = {7},
number = {CSCW2},
url = {https://doi.org/10.1145/3610192},
doi = {10.1145/3610192},
journal = {Proc. ACM Hum.-Comput. Interact.},
month = oct,
articleno = {343},
numpages = {26}
}

@inproceedings{Kim1,
author = {Kim, Soomin and Eun, Jinsu and Oh, Changhoon and Suh, Bongwon and Lee, Joonhwan},
title = {Bot in the Bunch: Facilitating Group Chat Discussion by Improving Efficiency and Participation with a Chatbot},
year = {2020},
isbn = {9781450367080},
publisher = {Association for Computing Machinery},
address = {New York, NY, USA},
url = {https://doi.org/10.1145/3313831.3376785},
doi = {10.1145/3313831.3376785},
pages = {1–13},
numpages = {13},
location = {Honolulu, HI, USA},
series = {CHI '20}
}

@article{Kim2,
author = {Kim, Soomin and Eun, Jinsu and Seering, Joseph and Lee, Joonhwan},
title = {Moderator Chatbot for Deliberative Discussion: Effects of Discussion Structure and Discussant Facilitation},
year = {2021},
issue_date = {April 2021},
publisher = {Association for Computing Machinery},
address = {New York, NY, USA},
volume = {5},
number = {CSCW1},
url = {https://doi.org/10.1145/3449161},
doi = {10.1145/3449161},
journal = {Proc. ACM Hum.-Comput. Interact.},
month = apr,
articleno = {87},
numpages = {26}
}

@article{Tessler,
author = {Michael Henry Tessler  and Michiel A. Bakker  and Daniel Jarrett  and Hannah Sheahan  and Martin J. Chadwick  and Raphael Koster  and Georgina Evans  and Lucy Campbell-Gillingham  and Tantum Collins  and David C. Parkes  and Matthew Botvinick  and Christopher Summerfield },
title = {AI can help humans find common ground in democratic deliberation},
journal = {Science},
volume = {386},
number = {6719},
pages = {eadq2852},
year = {2024},
doi = {10.1126/science.adq2852},
URL = {https://www.science.org/doi/abs/10.1126/science.adq2852},
eprint = {https://www.science.org/doi/pdf/10.1126/science.adq2852},
}

@misc{wiki:norms,
author = {Wikipedia},
title = {Wikipedia:Expectations and norms of the Wikipedia community},
year = {2024},
url = {https://en.wikipedia.org/wiki/Wikipedia:Expectations_and_norms_of_the_Wikipedia_community#Main_policies,_guidelines_and_essays},
lastaccessed ={Feb 09, 2026}
}

@misc{wiki:essay,
author = {Wikipedia},
title = {Wikipedia:Essays},
year = {2026},
url = {https://en.wikipedia.org/wiki/Wikipedia:Essays},
lastaccessed ={Feb 10, 2026}
}

@inproceedings{Cosley,
author = {Welser, Howard T. and Cosley, Dan and Kossinets, Gueorgi and Lin, Austin and Dokshin, Fedor and Gay, Geri and Smith, Marc},
title = {Finding social roles in Wikipedia},
year = {2011},
isbn = {9781450301213},
publisher = {Association for Computing Machinery},
address = {New York, NY, USA},
url = {https://doi.org/10.1145/1940761.1940778},
doi = {10.1145/1940761.1940778},
booktitle = {Proceedings of the 2011 IConference},
pages = {122–129},
numpages = {8},
location = {Seattle, Washington, USA},
series = {iConference '11}
}

@inproceedings{Pirolli&Card,
  title={The sensemaking process and leverage points for analyst technology as identified through cognitive task analysis},
  author={Pirolli, Peter and Card, Stuart},
  booktitle={Proceedings of international conference on intelligence analysis},
  volume={5},
  number={1},
  pages={2--4},
  year={2005},
  organization={McLean, VA, USA}
}

@article{weber1981some,
  title={Some categories of interpretive sociology},
  author={Weber, Max},
  journal={The Sociological Quarterly},
  volume={22},
  number={2},
  pages={151--180},
  year={1981},
  publisher={Taylor \& Francis}
}

@article{cho,
author = {Cho, Soobin and Zachry, Mark and McDonald, David W.},
title = {Towards Insider Summarization for Mediation Instead of Moderation: Examining Wikipedian Views on Key Elements of Discussion Summaries},
year = {2025},
issue_date = {November 2025},
publisher = {Association for Computing Machinery},
address = {New York, NY, USA},
volume = {9},
number = {7},
url = {https://doi.org/10.1145/3757681},
doi = {10.1145/3757681},
journal = {Proc. ACM Hum.-Comput. Interact.},
month = oct,
articleno = {CSCW500},
numpages = {25}
}
